\documentclass[preprint,aps,prd,showpacs,nofootinbib,floatfix]{revtex4}
\usepackage{graphicx}
\usepackage{amsmath}
\usepackage{amssymb}
\usepackage{bm}

\begin{document}

\title{\mbox{}\\[10pt]
Estimating the Production Rate \\ 
of Loosely-bound Hadronic Molecules \\ 
using Event Generators}
%%%%%%%%%%%%%%%%%%%%%%%%%%%%%%%%%%%%%%%%%%%%%%%%%%%%%%%%%%%%%%%%%%%%%%%%%%%%%%

\author{Pierre  Artoisenet and Eric Braaten}
%\thanks{}
\affiliation{Physics Department, Ohio State University, Columbus,
Ohio 43210, USA}

\date{\today}
%%%%%%%%%%%%%%%%%%%%%%%%%%%%%%%%%%%%%%%%%%%%%%%%%%%%%%%%%%%%%%%%%%%%%%%%%%%%%%
\begin{abstract}
We examine the use of hadronic event generators, such as Pythia or Herwig,
to estimate the production rate of  
loosely-bound hadronic molecules, such as the deuteron and the $X(3872)$.
In the case of the deuteron, we point out that there are large uncertainties 
in the normalization of the predictions using event generators, 
because baryon pair distributions are not among the inputs 
used to tune the event generators.  
Predictions using Pythia for anti-deuteron production in $\Upsilon$ 
decay are compared to measurements by the CLEO Collaboration. 
They suggest that Pythia overpredicts the probability of producing
pairs of baryons, at least in $\Upsilon$ decay into three gluons,
and that the standard value of the coalescence parameter underpredicts the
probability for formation of a deuteron from a neutron and proton 
with small relative momentum.
In the case of the $X(3872)$, we discuss a proposed upper bound 
on the prompt cross section at the Tevatron that
has been used as an argument against
the $X(3872)$ being a loosely-bound charm meson molecule.
We demonstrate that this proposed upper bound is invalid
by showing that the analogous upper bound for the anti-deuteron
would be smaller than the observed anti-deuteron cross section.
\end{abstract}

% 12.38.-t   Quantum chromodynamics
% 12.39.St  Factorization
% 13.20.Gd  Decays of J/psi, Upsilon, and other quarkonia
% 14.40.Gx   Mesons with S=C=B=0, mass > 2.5 GeV (including quarkonia)

%\keywords{}

\maketitle


\section{Introduction}
\label{sec:intro}

Quantum mechanics predicts that a bound state that is sufficiently 
close to a 2-body threshold and that couples to that threshold
through a short-range S-wave interaction
has universal properties that depend only on its binding energy.  
Such a bound state is necessarily a loosely-bound molecule in which the 
constituents are almost always separated by more than the range.  
One of the universal predictions is that the root-mean-square (rms) 
separation of the constituents is $(4 \mu E_X)^{-1/2}$, 
where $E_X$ is the binding energy of the resonance and $\mu$ is the 
reduced mass of the two constituents.  
As the binding energy is tuned to zero, the size of the molecule 
increases without bound.
A classic example of a loosely-bound S-wave 
molecule is the deuteron, which is a bound state of the proton 
and neutron with binding energy 2.2~MeV.  The proton and neutron are 
correctly predicted to have a large rms separation of about 3.1~fm.

An even more ideal example of a loosely-bound S-wave molecule is the 
charmonium-like state $X(3872)$, 
provided that its $J^{PC}$ quantum numbers are $1^{++}$.  
Measurements of its mass in the decay mode $J/\psi\, \pi^+ \pi^-$ 
indicate that it is below the threshold for $D^{*0} \bar D^0$ by 
$0.42 \pm 0.39$~MeV 
\cite{Abazov:2004kp,Aubert:2008gu,Belle:2008te,Aaltonen:2009vj}.  
If its quantum numbers are $1^{++}$, 
it has an S-wave coupling to $D^{*0} \bar D^0$.  In that case, it must be a 
loosely-bound molecule whose constituents are the superposition 
$D^{*0} \bar D^0 + D^0 \bar D^{*0}$.  The constituents are predicted 
to have a large rms separation of $4.9^{+13.4}_{~-1.4}$~fm.  

The production rate of a deuteron or anti-deuteron in high energy 
collisions is an important problem for several reasons.  
Anti-deuterons can be produced 
by the annihilation or decay of very massive dark-matter particles.  
Thus they provide a low-background channel for the indirect detection 
of dark matter \cite{Cui:2010ud}.  
The production of deuterons and anti-deuterons has been observed in 
relativistic heavy ion collisions \cite{Adler:2001uy,Adler:2004uy}.
Their production serves as a probe of the 
expanding and cooling hadronic fluid at the time of its freeze-out 
into free-streaming hadrons.
The production of an anti-deuteron has also been observed 
in many high energy physics experiments, 
including $\Upsilon$ decays \cite{Albrecht:1989ag,Asner:2006pw}, 
$p \bar p$ collisions \cite{Alexopoulos:2000jk},
photoproduction \cite{Aktas:2004pq},
$Z^0$ decays \cite{Schael:2006fd},
and deep inelastic electron scattering \cite{Chekanov:2007mv}.
To explain the production rate quantitatively in these experiments 
is a challenge.
The production rate of the $X(3872)$ is important for understanding 
the nature of some of the new $c \bar c$ mesons above the open charm 
threshold that have been discovered in recent years \cite{Godfrey:2008nc}.
Thus far, the $X(3872)$ has been observed only in decays of $B$ mesons and
through inclusive production in $p \bar p$ collisions. 
It has been claimed that the observed prompt production rate of the $X(3872)$ 
at the Tevatron is orders of magnitude too large to be 
compatible with its identification as a loosely-bound S-wave
molecule \cite{Bignamini:2009sk}.   A subsequent analysis 
challenged this conclusion \cite{Artoisenet:2009wk}.  The resolution 
of the controversy has important implications for studies 
of the $X(3872)$ in experiments at the Large Hadron Collider.

Estimating the production rate of a loosely-bound S-wave molecule 
in high energy collisions is also an interesting problem.  
Intuitively, one expects the cross section to be very small, 
because one would expect the binding of the constituents into a molecule 
to be easily disrupted by the enormous energies available 
in a high energy collision.  On the other hand,  
since the constituents of the molecule are almost always outside the range 
of their interactions, they must be subject to a very strong force 
during the small fraction of time in which they are close together.
This strong force also operates between constituents that are 
produced with small relative momentum in a high energy collision.
The production rate of the molecule involves the interplay 
between this very strong force and the very weak binding.  

One tool that can be helpful in estimating the production rate 
of a loosely-bound S-wave hadronic molecule is a hadronic event generator, 
such as Pythia \cite{Sjostrand:2006za} or Herwig \cite{Corcella:2002jc}.
These event generators can be interpreted as purely phenomenological models 
for hadron production with numerous parameters that have been adjusted 
to fit data from many high energy physics experiments.  
They should provide accurate predictions for observables that are 
sufficiently similar to the ones that have been used to tune the 
parameters, but one should be wary of applying them to new phenomena.  
They may be able to take into account 
the effects of generic hadronic interactions,  
but they should not be expected to take into account the effects 
of finely-tuned interactions, such as those responsible for the existence 
of loosely-bound hadronic molecules.  Event generators have been used to
estimate the production rate of anti-deuterons in the annihilation 
of dark-matter particles \cite{Kadastik:2009ts}.  They have also been
applied to the production rate of the $X(3872)$ in hadron colliders 
\cite{Bignamini:2009sk,Artoisenet:2009wk,Bignamini:2009fn}.

In this paper, we address some of the issues involved in using 
hadronic event generators to estimate the production rate of 
loosely-bound S-wave hadronic molecules.   
In Section~\ref{sec:EvGen}, we discuss the use of an
event generator to estimate the production rate of the anti-deuteron. 
In Section~\ref{sec:CLEO}, we compare measurements of 
anti-deuteron production in $\Upsilon$ decays by the CLEO Collaboration 
with predictions from an event generator. 
In Section~\ref{sec:Xprod}, we discuss the controversy involving the use of 
event generators to estimate the prompt production rate of the $X(3872)$.
We discuss our results in Section~\ref{sec:conc}.

\section{Event-Generator Model for Deuteron Production}
\label{sec:EvGen}

The {\it coalescence model} is a purely phenomenological model 
for deuteron and anti-deuteron production \cite{Csernai:1986qf}.
According to this model, the differential distribution for a deuteron of 
momentum $\bm{P}$ is the product of the differential distributions
for a neutron and a proton with equal momenta $\frac12 \bm{P}$
multiplied by a Lorentz boost factor $E/2m_N$
and by a phenomenological constant.
That constant is often expressed as the volume $4 \pi p_0^3/3$
of a sphere in momentum space.  
The coalescence model can be ``derived'' from two assumptions:
\begin{enumerate}
\item
A neutron and a proton will bind to form a deuteron if they 
are produced with relative momentum less than $p_0$.
\item
The joint probability distribution for producing $n$ and $p$ 
factors into the product of independent probabilities 
for $n$ and $p$.
\end{enumerate} 
From an analysis of data on anti-deuteron production in 
proton-proton and proton-nucleus collisions with nucleon-nucleon 
center-of-mass energies in the range 20 to 53~GeV,
the coalescence parameter has been determined to be
$p_0 = 79$~MeV \cite{Duperray:2005si}.
We will use $p_0 = 80$~MeV to avoid the implication 
that this parameter can be determined with two digits of accuracy.
We will refer to this value as the {\it standard coalescence parameter}
for the deuteron.

Kadastik, Raidal, and Strumia recently pointed out that the 
coalescence model fails dramatically for the production of 
anti-deuterons in the annihilation of a pair of heavy dark-matter 
particles \cite{Kadastik:2009ts}.  It predicts incorrectly that the 
probability for producing an anti-deuteron scales as $1/M^2$, 
where $M$ is the mass of the dark-matter particle.
However the probability is actually a slowly varying function of $M$.
The reason the coalescence model fails is that a pair of 
dark-matter particles annihilates predominantly into two jets,
and the $\bar d$ is almost always produced by 
the coalescence of $\bar n$ and $\bar p$ within the same jet.  
While the separate probability distributions 
for $\bar n$ and $\bar p$ are spherically symmetric, 
the joint probability distribution for $\bar n$ and $\bar p$ 
is sharply peaked for $\bar n$ and $\bar p$ in the same direction.
Thus assumption 2 of the coalescence model breaks down completely.

Kadastik, Raidal, and Strumia proposed an alternative model 
for the production of anti-deuterons that gives the correct scaling 
behavior when the production is dominated by jets \cite{Kadastik:2009ts}.  
They retained assumption 1, but assumption 2 was replaced by 
an alternative assumption:
\begin{enumerate}
\item[$2^{\, \prime}$.]
The joint probability distribution for producing $n$ and $p$ 
can be calculated using a hadronic event generator, 
such as Pythia or Herwig.
\end{enumerate} 
The model consisting of assumptions 1 and $2^{\, \prime}$
implies a simple equation for the inclusive deuteron cross section:
%-----------------
\begin{equation}
\sigma[d] = \sigma_\textrm{naive}[np(k < p_0)]. 
\label{sigma-cm}
\end{equation}
%-----------------
The subscript ``naive" on the right side refers to the 
$n p$ cross section being calculated using a method 
that is not informed about the 
fine-tuning of interactions that is
responsible for binding the $n$ and $p$ into $d$.  
In Ref.~\cite{Kadastik:2009ts}, the authors used this 
model to calculate the $\bar d$ yield per dark-matter 
annihilation event for various pairs of jets, using Pythia 
as their event generator.
For a dark-matter particle with a mass $M$ of about 100~GeV, 
the yields are larger than those predicted by the coalescence
model by more than an order of magnitude 
and the discrepancy increases like $M^2$.

It should be obvious from its formulation that this model
is a purely phenomenological model with no fundamental justification.  
However this model also has a practical problem in 
that it relies on Pythia or Herwig to give the 
distribution for pairs of baryons.  
Measurements of single-baryon momentum distributions 
in various high energy physics experiments have
been used to tune these event generators, but,
to the best of our knowledge, information about baryon pairs has  
not been used. 
Thus one should allow at least for an unknown normalizing factor 
$K_{np}$ in its predictions for $np$ pair distributions.
This can be expressed as an alternative to the assumption 2
of the event-generator model:
\begin{enumerate}
\item[$2^{\, \prime\prime}$.]
The joint probability distribution for producing $n$ and $p$ 
can be calculated using a hadronic event generator, 
such as Pythia or Herwig, up to a normalizing factor $K_{np}$.
\end{enumerate} 
The model consisting of assumptions 1 and $2^{\, \prime \prime}$
implies a simple equation for the inclusive deuteron cross section:
%-----------------
\begin{equation}
\sigma[d] = K_{np}~\sigma_\textrm{naive}[np(k < p_0)]. 
\label{sigma-egm}
\end{equation}
%-----------------
We will refer to this model as the {\it event-generator model}.

In the spirit of hadronic event generators,
the normalizing factor $K_{np}$ and the coalescence parameter $p_0$ 
should be treated as phenomenological parameters that must be 
determined from data.  Their values need not be the same in all 
high energy physics processes.  Their values for large 
transverse momentum processes, which are dominated by jets, 
could be different from their values for low transverse momentum processes.  
They could have different values for processes initiated by 
quarks and antiquarks than for processes initiated by gluons. 
In the absence of data that can be used to determine $p_0$ and $K_{np}$
separately, the most reliable predictions of 
the event-generator model will be for ratios of
observables in which $K_{np}$ cancels.

The ALEPH Collaboration has measured the inclusive 
decay rate of the $Z^0$ into
an anti-deuteron \cite{Schael:2006fd}.
The number of anti-deuterons per hadronic $Z^0$ decay is
%-----------------
\begin{equation}
\frac{{\cal B}[Z^0 \to \bar d + X]}
    {{\cal B}[Z^0 \to \textrm{hadrons}]}
= (5.9 \pm 1.8 \pm 0.5) \times 10^{-6}.  
\label{R:Z0}
\end{equation}
%-----------------
In Ref.~\cite{Kadastik:2009ts}, the production rate of $\bar d$
in $Z^0$ decay was calculated using the Pythia event generator.  
Taking the measurement in Eq.~(\ref{R:Z0}) as the input, the 
coalescence parameter was determined to be
$p_0 = 81 \pm 9~\textrm{MeV}$.
This is consistent to within errors with 
the standard value $p_0 = 80$~MeV.
In the event-generator model, the branching ratio in Eq.~(\ref{R:Z0})
is sensitive only to the combination $K_{np} p_0^3$.
Hadronic decays of the $Z^0$ are dominated by its decay into a
quark and antiquark, each of which hadronizes into a jet.
Thus a conservative conclusion from the calculation in 
Ref.~\cite{Kadastik:2009ts} 
is that $K_{np} p_0^3 \approx (80~\textrm{MeV})^3$
for $\bar d$ production in a jet initiated by a quark or antiquark.

\section{Anti-deuteron Production in $\bm{\Upsilon}$ Decays}
\label{sec:CLEO}

The high energy process for which there is the most information 
about anti-deuteron production is $\Upsilon$ decay.
In this section, we compare measurements of anti-deuteron production 
in $\Upsilon$ decay by the CLEO Collaboration \cite{Asner:2006pw}
with predictions of the event-generator model.

\subsection{CLEO measurements}

The CLEO Collaboration has studied the production 
of the deuteron and the anti-deuteron in a data sample of 
$2.2\times 10^7$ $\Upsilon(1S)$ decays \cite{Asner:2006pw}.  
The rates for the deuteron and
anti-deuteron are presumably equal, but the backgrounds are smaller 
for the anti-deuteron, because the CLEO detector is made of matter rather 
than antimatter.  They also studied the production of the anti-deuteron
$\bar d$ in $e^+e^-$ annihilation off the resonance.  In $e^+e^-$ annihilation, 
the production process is initiated by the decay of a virtual photon 
into a light quark-antiquark pair.  In $\Upsilon$ decay, 
the production process is initiated either by the annihilation of $b \bar b$ 
into a virtual photon, which then decays into a light $q \bar q$ pair, 
or by the direct annihilation of $b \bar b$ into partons, 
such as 3 gluons.  
The virtual-photon contributions to the inclusive partial width 
of $\Upsilon$ into $\bar d$ and to the total hadronic width 
of $\Upsilon$ can both be determined from measurements off the resonance.
CLEO therefore found it convenient to express their results 
in terms of the ``direct" branching fraction, in which the virtual-photon 
contributions have been subtracted from both the numerator and denominator.  
Their result for the direct branching fraction was 
 %-----------------
\begin{equation}
{\cal B}^\textrm{dir}[\Upsilon \to \bar d + X] 
= (3.36 \pm 0.23 \pm 0.25) \times 10^{-5}.  
\label{B:CLEO}
\end{equation}
%-----------------
One can interpret this as the inclusive branching fraction into $\bar d$ 
from the annihilation of $\Upsilon$ into 3 gluons.  
For $\bar d$ production from the decay of a virtual photon, 
CLEO set an upper bound on the inclusive 
branching fraction of about $10^{-5}$.  
Thus the production rate of $\bar d$ is significantly 
larger in gluon-initiated processes than in $q \bar q$-initiated processes.

The presence of the anti-deuteron in an $\Upsilon$ decay event 
implies that the event also includes at least two baryons.  
The CLEO Collaboration studied the nature of the associated baryons.
Their results were consistent with the  $\bar d$  being accompanied 
by $nn$, $np$, and $pp$ with probabilities 25\%, 50\%, and 25\%, respectively.  
They also found 3 events out of their 338 $\bar d$ candidates 
in which the $\bar d$ was accompanied by a $d$.  
The ratio of these numbers of events provides an estimate of the 
branching ratio for inclusive $\bar d + d$ and inclusive $\bar d$:
%-----------------
\begin{equation}
\frac{{\cal B}^\textrm{dir}[\Upsilon \to \bar d + d + X]} 
    {{\cal B}^\textrm{dir}[\Upsilon \to \bar d + X]} 
\approx 0.009.  
\label{BB:CLEO}
\end{equation}
%-----------------
The naive assumption that $N$ events can have fluctuation of 
$\pm \sqrt{N}$ implies that the error bar is at least as large as 
$\pm 0.006$. 

\subsection{Event-generator model}
\label{sec:Pythia}

%%%%%%%%%%%%%%%%%%%%%%%%%%%%%%%%%%%%%%%%%%%%%%%%%
\begin{figure}[t]
\centerline{\includegraphics*[height=8cm,angle=0,clip=true]{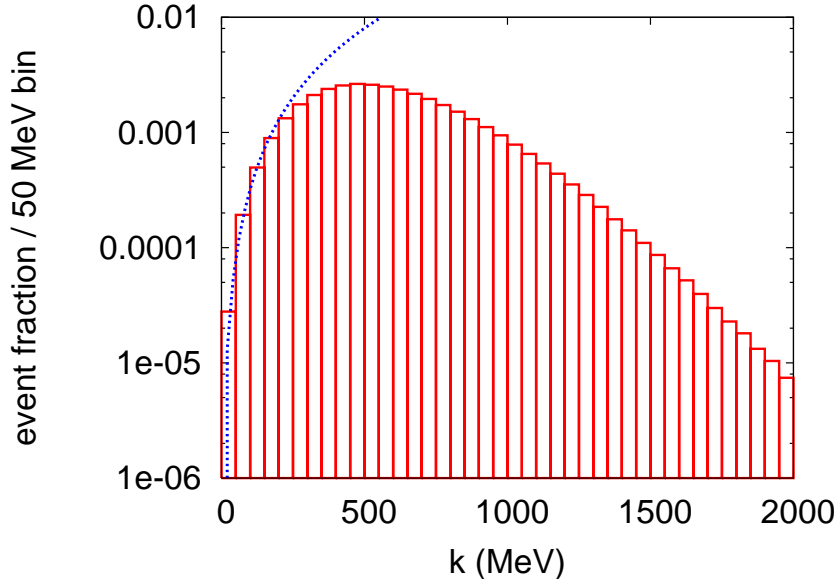}}
\vspace*{0.0cm}
\caption{Fraction of $\Upsilon \rightarrow ggg$ events generated by Pythia 
with an $\bar n \bar p$ pair in the final state as a function
of the relative momentum $k$ between the $\bar n$ and $\bar p$.
The dotted line is a phase space distribution proportional to $k^2$.
\label{krel_nbar_pbar}}
\end{figure}
%%%%%%%%%%%%%%%%%%%%%%%%%%%%%%%%%%%%%%%%%%%%%%%%%%

The event-generator model can be used to predict 
the production rate of an anti-deuteron
from the annihilation of $\Upsilon$ into 3 gluons. 
We have generated $140 \times 10^{6}$
$\Upsilon \rightarrow ggg$ events using Pythia. 
The fraction of $\Upsilon \rightarrow ggg$ events 
that include an $\bar{n} \bar{p}$ pair
is displayed in Figure~\ref{krel_nbar_pbar} as a function 
of the relative momentum $k$ between the $\bar{n}$ and $\bar{p}$.
The fraction of events follows a phase space distribution 
proportional to $k^2$ out to about $200$~MeV.
We can therefore use the phase space distribution 
to calculate the fraction of $\bar{n} \bar{p}$ events 
with $k<p_0$.
The prediction of the event-generator model for the
direct branching fraction into $\bar d$ is
%-----------------
\begin{equation}
{\cal B}^\textrm{dir}[\Upsilon \to \bar n \bar p (k < p_0) + X]
= 1.1 \times 10^{-4}~K_{np} \left( \frac{p_0}{80~\textrm{MeV}} \right)^3.
\label{B:Pythia}
\end{equation}
%-----------------
If we set $K_{np} = 1$ and $p_0 = 80$~MeV,
this prediction is larger than the CLEO measurement
of the direct branching fraction in Eq.~(\ref{B:CLEO})
by about a factor of 3.5.

%%%%%%%%%%%%%%%%%%%%%%%%%%%%%%%%%%%%%%%%%%%%%%%%%
\begin{figure}[t]
\centerline{\includegraphics*[height=8cm,angle=0,clip=true]{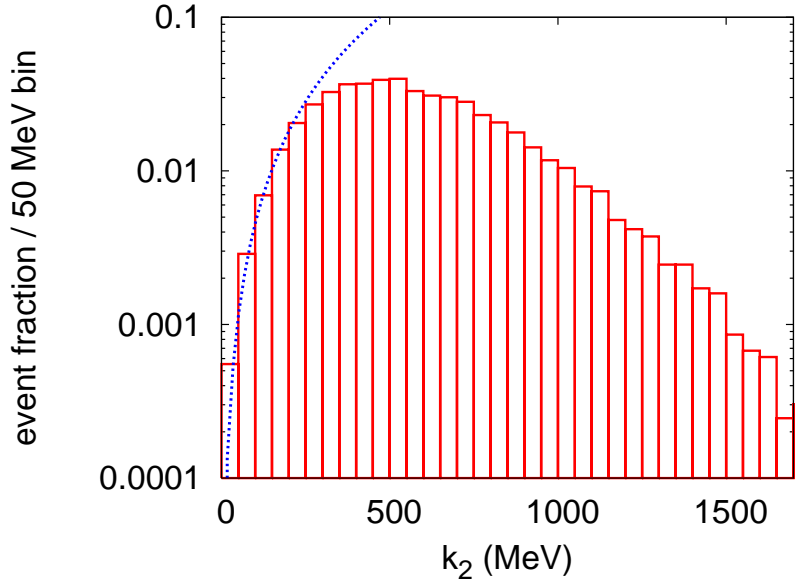}}
\vspace*{0.0cm}
\caption{Fraction of 
$\Upsilon \rightarrow ggg \rightarrow 
\bar{n} \bar{p}(k_1 < 80~\textrm{MeV}) + X$ 
events generated by Pythia with an $n p$ pair in the final state
as a function
of the relative momentum $k_2$ between the $n$ and $p$.
The dotted line is a phase space distribution proportional to $k_2^2$.
\label{krel_np}}
\end{figure}
%%%%%%%%%%%%%%%%%%%%%%%%%%%%%%%%%%%%%%%%%%%%%%%%%%

The production of inclusive $\bar d + d$ events can be studied in the 
event-generator model by counting the events
with both an $ \bar n \bar p$ pair and an  $n p$ pair, 
each of which has relative momentum smaller than $p_0$. 
The fraction of 
$\Upsilon \rightarrow ggg \rightarrow \bar{n} \bar{p}(k_1<p_0)+X$ events
that include an $np$ pair in the final state as a function
of the relative momentum $k_2$ between $n$ and $p$
is shown in Fig.~\ref{krel_np}.  
The fraction of events follows a phase space distribution 
proportional to $k_2^2$ out to about $200$~MeV.
We can therefore use the phase space distribution 
to calculate the fraction of $n p$ events 
with $k_2<p_0$.
The prediction of the event-generator model for the
ratio of the direct branching fractions into $\bar d + d$ and $\bar d$ is
%-----------------
\begin{equation}
\frac{{\cal B}^\textrm{dir}[\Upsilon \to \bar n \bar p (k_1 < p_0)
                                        + n p (k_2 < p_0) + X]} 
    {{\cal B}^\textrm{dir}[\Upsilon \to \bar n \bar p (k_1 < p_0) + X]} 
= 1.6 \times 10^{-3} \left( \frac{p_0}{80~\textrm{MeV}} \right)^3.
\label{BB:Pythia}
\end{equation}
%-----------------
The numerator is proportional to a single factor of $K_{np}$,
because the presence of the $\bar n \bar p$ pair 
requires an accompanying antibaryon pair.
Thus the normalizing factor $K_{np}$ cancels between the 
numerator and denominator.
If we set $p_0 = 80$~MeV in Eq.~(\ref{BB:Pythia}),
this prediction is smaller than the estimate
of the branching ratio from CLEO data in Eq.~(\ref{BB:CLEO})
by about a factor of 6.

The parameters $K_{np}$ and $p_0$ of the event-generator model
can be adjusted so that the predictions of the model 
in Eqs.~(\ref{B:Pythia}) and (\ref{BB:Pythia}) agree with the 
CLEO results in Eqs.~(\ref{B:CLEO}) and (\ref{BB:CLEO}).
Setting Eqs.~(\ref{B:Pythia}) and (\ref{B:CLEO}) equal, we get
$K_{np} p_0^3 = ( 53 \pm 5~\textrm{MeV} )^3$.
Setting Eqs.~(\ref{BB:Pythia}) and (\ref{BB:CLEO}) equal, we get
the estimate $p_0 \approx 140$~MeV.  
Allowing for a statistical error of $\pm 1/\sqrt{3}$ 
in Eq.~(\ref{BB:CLEO}), 
the estimate for $p_0$ ranges from 105~MeV to 163~MeV.
Combining the two results,
we obtain the estimate $K_{np} \approx 0.05$ with 
an error that is at least $\pm 0.03$.
The large errors in our estimates for $p_0$
and $K_{np}$ come from the small number of $\bar d + d$
candidates observed in the experiment.
It is somewhat surprising that $K_{np}$ is 
one or two orders of magnitude smaller than 1.
Pythia predicts that 3.6\% of the $\Upsilon \to ggg$ events 
include $np$ and therefore also two antibaryons.
In these events, almost half the 9.46~GeV of available energy 
goes into the rest energy of the four baryons and antibaryons.
The predictions of an event generator for rare events like these
can be expected to have large errors unless 
they are tuned to data.  Since double baryon production was not
used in the tuning of Pythia, it is plausible that there is a
large error in its prediction for inclusive $n p$ production.
If Pythia significantly overpredicts the probability 
of creating an $n p$ pair,
the standard coalescence parameter $p_0 = 80$~MeV must also 
underpredict the probability of their binding to form $d$,
at least in the process $\Upsilon \to ggg$.

\section{The $\bm{X(3872)}$ Production Controversy}
\label{sec:Xprod}

Hadronic event generators have been used by two different groups
to estimate the production rate of the $X(3872)$.
Their estimates differ by orders of magnitude and lead to opposite
conclusions about whether the $X(3872)$ can be a loosely-bound charm meson
molecule.  In this section, we present a critical evaluation of 
those estimates.

\subsection{Estimates of the $\bm{X(3872)}$ production rate}

The quantum numbers of the $X(3872)$ have been narrowed
down experimentally to two possibilities,  $1^{++}$ or $2^{-+}$,
by the observation of its decay into $J/\psi \gamma$ \cite{Abe:2005ix}
and by an analysis of its decays into 
$J/\psi\, \pi^+ \pi^-$  \cite{Abulencia:2006ma}.  
The observation of its decay into $D^{0} \bar D^0 \pi^0$ \cite{Gokhroo:2006bt},
whose threshold is lower by only about 7~MeV, 
disfavors spin 2 because of angular momentum suppression.  
On the other hand, a recent analysis of decays into 
$J/\psi\, \pi^+ \pi^-\pi^0$ favors negative parity \cite{delAmoSanchez:2010jr}.  
Thus whether the quantum numbers of the $X(3872)$ 
are $1^{++}$ or $2^{-+}$ remains an open experimental question.  
We will assume that they are $1^{++}$,
in which case the $X(3872)$ must be a loosely-bound charm meson molecule
whose particle content is
%-----------------
\begin{equation}
X = \frac{1}{\sqrt2} \left( D^{*0} \bar D^0 + D^{0} \bar D^{*0} \right). 
\label{X-DbarD}
\end{equation}
%-----------------
Thus, if the event-generator model can be used to calculate 
the production rate of loosely-bound hadronic molecules, 
it should be applicable to the $X(3872)$.

The production of $X(3872)$ in high energy hadron collisions 
comes from two mechanisms:
the production of $b$ hadrons followed by their weak decay into $X(3872)$ 
and the prompt production of $X(3872)$ through QCD mechanisms.
The prompt cross section for $X(3872)$ at the Tevatron can be estimated 
from measurements by the CDF Collaboration \cite{Bauer:2004bc}.
The cross section for $X(3872)$ 
with transverse momentum $p_T > 5$~GeV and rapidity $|y| < 0.6$ is
\cite{Bignamini:2009sk,Artoisenet:2009wk}
%-----------------
\begin{equation}
\sigma[X(3872)]~\textrm{Br}[X \to J/\psi\, \pi^+ \pi^-] 
= 3.1 \pm 0.7~\textrm{nb}, 
\label{sigma-CDF}
\end{equation}
%-----------------
up to corrections for acceptances and efficiencies
that are expected to be small.
From measurements of decays of $X(3872)$ produced in $B$ meson decays, 
one can infer that the branching fraction for $X(3872)$ to decay 
into $J/\psi\, \pi^+ \pi^-$ is less than about 10\% \cite{Artoisenet:2009wk}. 
Thus the experimental lower bound on the cross section
for $X(3872)$ is about 30~nb.

Two groups have used event generators to estimate the prompt 
cross section for the $X(3872)$ at the 
Tevatron $p \bar p$ collider \cite{Bignamini:2009sk,Artoisenet:2009wk}.
Both estimates are expressed in terms of naive cross sections for
the inclusive production of $D^{*0} \bar D^0$ and $D^{0} \bar D^{*0}$ 
with relative momentum $k$
integrated up to some maximum $k_\textrm{max}$.
Hadronic event generators, such as Pythia or Herwig, can be used
to calculate the naive cross sections for the charm meson pairs.
These event generators are tuned to 
reproduce charm meson distributions in various high energy experiments,
but they have not been tuned to reproduce charm meson pair 
distributions.  Thus one should allow at least for an unknown 
normalizing factor $K_{D^* \bar D}$ in their predictions for 
charm meson pair distributions.

The dramatic discrepancy between the estimates in
Refs.~\cite{Bignamini:2009sk,Artoisenet:2009wk} does not depend 
on the event generators.
In Ref.~\cite{Bignamini:2009sk} (BGPPS), the authors proposed an upper 
bound on the prompt cross section for the $X(3872)$:
%-----------------
\begin{equation}
\sigma[X(3872)] < \frac12 K_{D^* \bar D}
\left( \sigma_\textrm{naive}[D^{*0} \bar D^0(k < k_\textrm{max})]
+ \sigma_\textrm{naive}[D^{0} \bar D^{*0}(k < k_\textrm{max})] \right) . 
\label{sigma-ubX}
\end{equation}
%-----------------
Their prescription for $k_\textrm{max}$ was proportional to the 
binding momentum $\gamma_X = \sqrt{2 \mu E_X}$ of the $X(3872)$,
where $\mu$ is the reduced mass of $D^{*0} \bar D^0$.
In Ref.~\cite{Artoisenet:2009wk} (AB), the authors proposed an 
order-of-magnitude estimate for the prompt cross section for the $X(3872)$:
%-----------------
\begin{equation}
\sigma[X(3872)] \approx \frac{3 \pi \gamma_X}{ k_\textrm{max}} K_{D^* \bar D}
\left( \sigma_\textrm{naive}[D^{*0} \bar D^0(k < k_\textrm{max})]
+ \sigma_\textrm{naive}[D^{0} \bar D^{*0}(k < k_\textrm{max})] \right). 
\label{sigma-estX}
\end{equation}
%-----------------
Their prescription for $k_\textrm{max}$ was the inverse of the 
range of the interactions between the charm mesons,
give or take a factor of 2. 
Taking $1/m_\pi$ as an estimate of the range,
their prescription reduced to $k_\textrm{max} = m_\pi$, 
give or take a factor of 2. 
Now the naive cross sections in
Eqs.~(\ref{sigma-ubX}) and (\ref{sigma-estX}) scale like 
$k_\textrm{max}^3$ from phase space.
The ratio of the estimate in Eq.~(\ref{sigma-estX}) with 
$k_\textrm{max} = m_\pi$ to the proposed upper bound in 
Eq.~(\ref{sigma-ubX}) with $k_\textrm{max} = \gamma_X$
is therefore $6 \pi (m_\pi/\gamma_X)^2$.
For $E_X = 0.4$~MeV, this ratio is about 530.  
Thus the estimate in Eq.~(\ref{sigma-estX}) 
is more than two orders of magnitude larger than the proposed upper bound in 
Eq.~(\ref{sigma-ubX}).  This dramatic discrepancy implies 
that there must be a serious conceptual error in the 
derivation of either the upper bound in Eq.~(\ref{sigma-ubX})
or the estimate in Eq.~(\ref{sigma-estX}) or both. 
 
In Ref.~\cite{Bignamini:2009sk}, BGPPS used both 
Pythia and Herwig to calculate the upper bound in Eq.~(\ref{sigma-ubX})
for the prompt $X(3872)$ cross section at the Tevatron.  
They used measurements of $D^0 D^{*-}$ production at the Tevatron
to determine the normalizing factor $K_{D^* \bar D}$. 
The upper bounds on $\sigma[X]$ calculated by BGPPS 
using $k_\textrm{max} = 35$~MeV
were 0.11~nb using Pythia and 0.07~nb using Herwig.
These theoretical upper bounds 
are more than two orders of magnitude smaller than the experimental 
lower bound of about 30~nb implied by Eq.~(\ref{sigma-CDF}).
BGPPS concluded that the 
$X(3872)$ was unlikely to be a loosely-bound charm meson molecule.  

This conclusion was challenged in Ref.~\cite{Artoisenet:2009wk}.  
AB pointed out that the constituents of a loosely-bound S-wave molecule
need not be created with relative momentum 
of order the binding momentum $\gamma_X$.
Rescattering of the constituents allows the formation of a bound state from
constituents that are created with much larger relative momentum. 
They argued that a more appropriate value for $k_\textrm{max}$
in the upper bound in Eq.~(\ref{sigma-ubX}) is the inverse of
the effective range for the charm mesons. 
The effective range is not known, 
but a reasonable order-of-magnitude estimate is $1/m_\pi$.
If the upper limit $k_\textrm{max} = 35$~MeV used in 
Ref.~\cite{Bignamini:2009sk} is replaced by $k_\textrm{max} = m_\pi$, 
the upper bound is increased by about a factor of 60.
This removes much of the discrepancy between the upper bound 
in Eq.~(\ref{sigma-ubX}) and the experimental 
lower bound implied by Eq.~(\ref{sigma-CDF}).

In Ref.~\cite{Artoisenet:2009wk}, AB used 
Pythia to calculate the estimate in Eq.~(\ref{sigma-estX}) for the 
prompt $X(3872)$ cross section at the Tevatron. 
They also used Madgraph to generate the Monte Carlo events more
efficiently.  They followed Ref.~\cite{Bignamini:2009sk} 
in using measurements of $D^0 D^{*-}$ production at the Tevatron
to determine the normalizing factor $K_{D^* \bar D}$.
The required factor ranges from 
$0.7$ to $1.6$ depending on the specific 
data used to determine the normalization.
For $E_X = 0.3$~MeV and $k_\textrm{max} = m_\pi$, 
they obtained the estimate $\sigma[X] \approx 6$~nb.
The experimental lower bound of about 30~nb implied by Eq.~(\ref{sigma-CDF})
can be accomodated by choosing $k_\textrm{max} > 300$~MeV.
Given the large uncertainties, AB concluded that the
observed prompt production rate of the $X(3872)$ at the Tevatron is  
compatible with its identification as a loosely-bound charm meson molecule.  

The dramatic difference in the conclusions of
Refs.~\cite{Bignamini:2009sk} and \cite{Artoisenet:2009wk}
concerning the nature of the $X(3872)$ comes from 
the dramatic conflict between the  upper bound in
Eq.~(\ref{sigma-ubX}) and the estimate in Eq.~(\ref{sigma-estX}).
We proceed to reexamine the derivation of these results.
For simplicity, we carry out the discussion in the specific
context of the deuteron.  This avoids the notational complexity 
associated with constituents of the $X(3872)$ being 
the superposition of charm mesons given in Eq.~(\ref{X-DbarD}).

\subsection{Upper bound of Ref.~\cite{Bignamini:2009sk} applied to the deuteron}

We first consider the upper bound in Eq.~(\ref{sigma-ubX}), 
which was derived by BGPPS in Ref.~\cite{Bignamini:2009sk}. 
The analogous upper bound for the inclusive production of the deuteron is
%-----------------
\begin{equation}
\sigma[d] < K_{np}~\sigma_\textrm{naive}[np(k < k_\textrm{max})].
\label{sigma-ub}
\end{equation}
%-----------------
The prescription of BGPPS for $k_\textrm{max}$ will be described below.
Their derivation of this upper bound begins by
expressing the inclusive cross section as the square 
of the production amplitude, summed over additional particles 
in the final state.  The production amplitude is approximated by
the product of the momentum-space wavefunction $\psi(k)$ for 
the deuteron and the production amplitude for an $np$ pair
with relative momentum $k$, integrated over the vector $\bm{k}$.
The range of the integral over $\bm{k}$ can be restricted to the region 
$0 < k < k_\textrm{max}$ in which the integrand has significant support.
By applying the Schwartz inequality to the square of the 
production amplitude, one can derive the inequality
%-----------------
\begin{equation}
\sigma[d] \le \sigma[np(k < k_\textrm{max})]~
\int \frac{d^3k}{(2 \pi)^3} |\psi(k)|^2~\theta(k < k_\textrm{max}).
\label{sigma-ubpsi}
\end{equation}
%-----------------
The last factor is the incomplete normalization integral for the 
wavefunction of the molecule, so it is less than 1. 
If the cross section $\sigma$ for $np$ with $k < k_\textrm{max}$
is dominated by generic hadronic scattering processes,
it can be approximated by a naive cross section $\sigma_\textrm{naive}$
that is not informed about the binding mechanism for the molecule.
It $\sigma_\textrm{naive}$ is calculated using an event generator, 
one should also allow for a normalization factor $K_{np}$ 
for the production of a pair of baryons.
This gives the upper bound in Eq.~(\ref{sigma-ub}).  

While the derivation of the upper bound in Eq.~(\ref{sigma-ub}) 
is plausible, its validity hinges on the value of $k_\textrm{max}$.
The right side of Eq.~(\ref{sigma-ub}) is a strictly increasing 
function of $k_\textrm{max}$, so the inequality is certainly satisfied 
for sufficiently large $k_\textrm{max}$.  The issue is whether
the prescription for $k_\textrm{max}$ used by BGPPS is valid
for a loosely-bound molecule.
Their prescription
was not stated clearly in Ref.~\cite{Bignamini:2009sk},
but a partial clarification is given in Ref.~\cite{Drenska:2010kg}.
It can be expressed as 
$k_\textrm{max} = k_0 + \Delta k$, where $k_0$ and $\Delta k$
are the typical momentum and the momentum spread in the bound state.
Their prescription for $k_0$ seems to be the binding momentum: 
$k_0 = \gamma_d \equiv \sqrt{m_N E_d}$.
Their prescription for $\Delta k$ seems to be the minimum spread 
in the momentum that is allowed by the uncertainty principle 
for a wavefunction whose rms separation is $\gamma_d^{-1}$:
$\Delta k = \gamma_d/2$.  
(The universal prediction for the rms separation 
in a loosely-bound molecule is $\gamma_d^{-1}/\sqrt2$.)
Since both $k_0$ and $\Delta k$ are 
proportional to $\gamma_d$, we can summarize their prescription by 
$k_\textrm{max} = 1.5~\gamma_d$.  
We proceed to critically examine this prescription.

The prescription $k_\textrm{max} = k_0 + \Delta k$
in Ref.~\cite{Bignamini:2009sk} is completely arbitrary.
One could equally well have used the prescription 
$k_\textrm{max} = a k_0 + b \Delta k$, where $a$ and $b$ are numerical
coefficients that are not too much larger than 1. 
This is important, because the naive cross section in 
Eq.~(\ref{sigma-ub}) scales like $k_\textrm{max}^3$ and is therefore
very sensitive to $k_\textrm{max}$.  A factor of 2 change in 
$k_\textrm{max}$ will change the upper bound by almost an order of magnitude.

The prescriptions for $k_0$ and $\Delta k$ used in 
Ref.~\cite{Bignamini:2009sk}, which are both proportional to $\gamma_d$,  
are not only arbitrary but they are physically incorrect.
More natural choices would have been the mean momentum $\bar k$
and the standard deviation $\Delta k$ for a loosely-bound molecule
with binding momentum $\gamma_d$.  The universal wavefunction 
in momentum space for such a molecule is
%-----------------
\begin{equation}
\psi(k) =
\frac{\sqrt{\gamma_d}}{\pi (k^2 + \gamma_d^2)}.
\label{psi-k}
\end{equation}
%-----------------
With this wavefunction, $\bar k$ is logarithmically ultraviolet divergent
and $\Delta k$ is linearly ultraviolet divergent.
The ultraviolet divergences are cut off by the range of the interaction 
between the constituents.  In the case of the deuteron, 
an appropriate choice for the range is the effective range $r_t = 1.76$~fm
for $n p$ scattering in the spin-triplet channel.
The physical interpretation of the divergences is that $\bar k$ is 
proportional to $\gamma_d$, with a coefficient that scales as 
$\log(1/\gamma_d r_t)$, and that $\Delta k$ scales as $1/r_t$.

That the upper bound in Eq.~(\ref{sigma-ub})
with the prescription $k_\textrm{max} = 1.5~\gamma_d$
is not valid
can also be demonstrated on phenomenological grounds.
We can regard Eq.~(\ref{sigma-cm}) with $p_0 = 80$~MeV
as an empirical deuteron cross section determined from the analysis
in Ref.~\cite{Duperray:2005si}.
The binding momentum of the deuteron is
$\gamma_d = 46$~MeV, so $1.5~\gamma_d \approx 70$~MeV.  
Since $k_\textrm{max} = 70$~MeV is smaller than $p_0 = 80$~MeV,
the proposed upper bound is smaller than the 
empirical deuteron cross section.
A more plausible choice for the upper limit $k_\textrm{max}$ 
in Eq.~(\ref{sigma-ub}) is $1/r_{t} \approx 110$~MeV.  
If we set $k_\textrm{max} = 110$~MeV, the upper bound
is larger than the empirical deuteron cross section in 
Eq.~(\ref{sigma-cm}) by about a factor of 2.6.

\subsection{Estimate of Ref.~\cite{Artoisenet:2009wk} applied to the deuteron}

We next consider the estimate in Eq.~(\ref{sigma-estX}), 
which was derived by AB in Ref.~\cite{Artoisenet:2009wk}. 
The analogous order-of-magnitude estimate for the case of the deuteron is
%-----------------
\begin{equation}
\sigma[d] \approx \frac{3}{4} 
\left( \frac{3 \pi \gamma_d}{k_\textrm{max}} \right) 
K_{np}~\sigma_\textrm{naive}[np(k < k_\textrm{max})], 
\label{sigma-est}
\end{equation}
%-----------------
where $k_\textrm{max}=1/r_t$, give or take a factor of 2.
This estimate is based on a rigorous relation between the 
cross section for a loosely-bound S-wave molecule 
and the cross section for its constituents that follows from the 
Migdal-Watson theorem \cite{Migdal-Watson}.
According to the Migdal-Watson theorem, the production amplitude for the 
constituents can be expressed as the product of their scattering amplitude
$(\gamma_d + i k)^{-1}$, where $\gamma_d$ is the binding momentum 
of the molecule, and a slowly varying function of the relative
momentum $k$ that depends on the short-distance details  
of the production process.  The production amplitude for the molecule 
has the same short-distance factor.
Eliminating the short-distance factor, we obtain a rigorous relation 
between the cross sections for the molecule and its constituents.
In the case of the deuteron, the relevant $np$ scattering channel 
is the $^3S_1$ channel and the relation is
%-----------------
\begin{equation}
\frac{d \sigma}{dk}[np(^3S_1, k)] =
\frac{k^2}{\pi \gamma_d (k^2 + \gamma_d^2)}\sigma[d] .
\label{dsigma-rig}
\end{equation}
%-----------------
If we integrate over the relative momentum up to $k_\textrm{max}$,
the relation becomes
%-----------------
\begin{equation}
\sigma[d] = 
\frac{\pi \gamma_d}
    {k_\textrm{max} - \gamma_d \arctan(k_\textrm{max}/\gamma_d)}
\sigma[np(^3S_1, k < k_\textrm{max})]. 
\label{sigma-rig}
\end{equation}
%-----------------
This rigorous relation holds for any $k_\textrm{max}$ 
in the region $k_\textrm{max} \ll 1/r_t$, where $r_t$ is the 
S-wave effective range, up to corrections suppressed by $k_\textrm{max} r_t$. 
Eq.~(\ref{sigma-rig}) implies that the cross section for $d$ 
is equal to that for $np$ if $k_\textrm{max} = 4.5~\gamma_d$:
%-----------------
\begin{equation}
\sigma[d] = 
\sigma[np(^3S_1, k < 4.5~\gamma_d)]. 
\label{sigma-equal}
\end{equation}
%-----------------
This relation does not apply to the deuteron, because 
the condition $4.5~\gamma_d \ll 1/r_t$ is violated.  
However the analogous relation might 
apply to more weakly bound molecules, such as the $X(3872)$. 

If we take the limit $\gamma_d \to 0$ in 
the rigorous relation in Eq.~(\ref{sigma-rig}), 
we see that the deuteron cross section decreases to 0 as $E_d^{1/2}$
as its binding energy decreases to 0.
This agrees with the conventional wisdom that the cross section for a
loosely-bound molecule should go to 0 as its binding energy goes to 0.
However naive phase space considerations suggest that 
the cross section should decrease as $E_d^{3/2}$.
For example, in the coalescence model, the order-of-magnitude of the 
coalescence parameter $p_0$ is often estimated by assuming 
that it is proportional to the binding momentum $\gamma_d$,
which would imply that the cross section decreases as $E_d^{3/2}$.
The actual suppression factor $E_d^{1/2}$ is much milder 
than the naive suppression factor $E_d^{3/2}$.

In Ref.~\cite{Artoisenet:2009wk}, 
AB used the rigorous relation in Eq.~(\ref{sigma-rig}) 
to obtain an order-of-magnitude estimate of the cross section
for a loosely-bound molecule. 
They chose $k_\textrm{max}$ to be the scale of the relative momentum $k$
at which the universal differential cross section $d\sigma/dk$, 
which approaches $\sigma[d]/\pi \gamma_d$ at large $k$,
becomes comparable to the naive differential cross section,
which scales as $k^2$ for small $k$.
The resulting estimate for $\sigma[d]$ is given
in Eq.~(\ref{sigma-est}).
The factor of 3/4 accounts for 3 of the 4 spin states of 
$np$ being in the spin-triplet channel in which there is binding. 
Since the naive cross section in Eq.~(\ref{sigma-est}) scales
like $k_\textrm{max}^3$, the estimate for $\sigma[d]$ 
is proportional to $k_\textrm{max}^2$.  An estimate of the momentum 
$k_\textrm{max}$ at which $d\sigma/dk$ becomes comparable to the 
$d\sigma_\textrm{naive}/dk$ is required to complete the estimate of 
$\sigma[d]$.

As an estimate of $k_\textrm{max}$, AB proposed 
the reciprocal of the effective range, give or take a factor of two.
In the case of the deuteron, the central estimate 
would be $1/r_{t} \approx 110$~MeV. 
Comparing with the phenomenological estimate in Eq.~(\ref{sigma-cm}), 
we see that this would correspond to a coalescence parameter 
$p_0 =(9 \pi \gamma_d/4 r_t^2)^{1/3} \approx 160$~MeV.  
Varying $k_\textrm{max}$ by a factor 2,
this theoretical estimate of the coalescence parameter $p_0$
varies from 100~MeV to 250~MeV.  This estimate is larger 
than the standard value 80~MeV obtained in 
Ref.~\cite{Duperray:2005si}.  It is interesting to note that the 
estimate of $p_0$ obtained from data on $\Upsilon$ decays
in Section~\ref{sec:Pythia} is also larger than the standard value.

\subsection{Hadronic activity}

In Ref.~\cite{Bignamini:2009fn}, the authors raised an issue 
concerning hadronic activity near a loosely-bound molecule.
If additional hadrons are produced that have small momentum 
relative to the molecule, their interactions with the constituents 
of the molecule can complicate the production process.
The order-of-magnitude estimate in Eq.~(\ref{sigma-est}),
which was based on the Migdal-Watson theorem, 
did not take into account the possibility 
of additional hadrons with small relative momentum.
The authors suggested that this cast doubts on the applicability of the 
Migdal-Watson theorem to an estimate of the production rate
in cases where there is significant hadronic activity near the molecule. 

%%%%%%%%%%%%%%%%%%%%%%%%%%%%%%%%%%%%%%%%%%%%%%%%%
\begin{figure}[t]
\centerline{\includegraphics*[height=8cm,angle=0,clip=true]{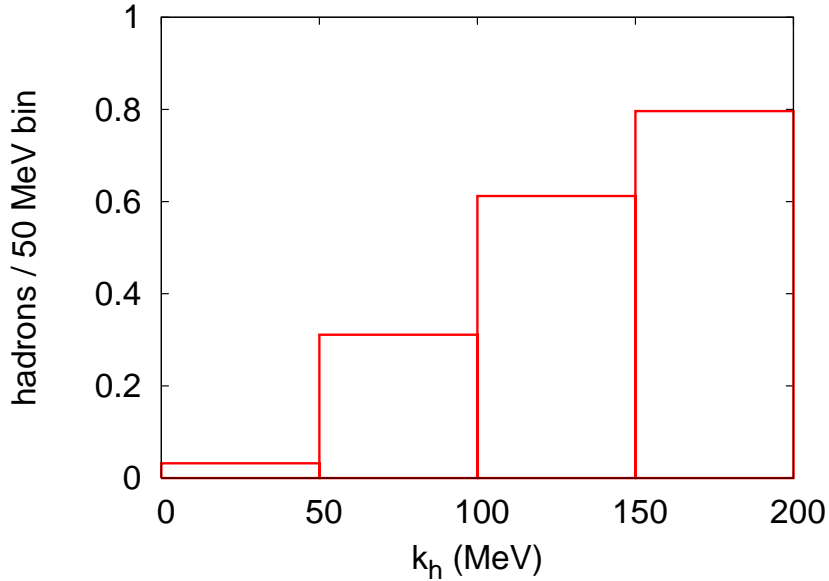}}
\vspace*{0.0cm}
\caption{Average number of hadrons in 
$\Upsilon \rightarrow ggg \rightarrow 
\bar{n} \bar{p}(k_1 < 80~\textrm{MeV}) + X$ events generated by Pythia
with respect to $k_h$, the smaller of the relative momenta 
of the hadron $h$ with respect to $\bar n$ and $\bar p$.}
\label{Hadr_activity}
\end{figure}
%%%%%%%%%%%%%%%%%%%%%%%%%%%%%%%%%%%%%%%%%%%%%%%%%%

In the case of anti-deuteron production in $\Upsilon$ decays, 
the only experimental information on the hadronic activity 
is that in the single $\bar d + d$ candidate event that was displayed
in Ref.~\cite{Asner:2006pw}, the $\bar d$ and $d$
were accompanied by 6 charged pions.
One can use an event generator to predict the hadronic activity.
In Fig.~\ref{Hadr_activity}, we show the prediction of Pythia for the 
number of additional hadrons $h$ produced in the events 
$\Upsilon \rightarrow ggg \rightarrow \bar{n} \bar{p}(k_1<p_0)+X$
as a function of the smaller of the relative momenta 
of the hadron $h$ with respect to $\bar n$ and $\bar p$, 
which we denote by $k_h$.
The average number of additional hadrons with $k_h < 100$~MeV is about 0.3.
More than $60\%$ of the events have no such additional hadron. 
Thus hadronic activity near the anti-deuteron does not seem to be 
a serious complication in $\Upsilon \rightarrow ggg$.

In the case of the production of the $X(3872)$ at the Tevatron,
the hadronic activity is larger because the Tevatron is a 
high-energy hadron collider.
In Ref.~\cite{Bignamini:2009fn}, 
event generators were used to predict the hadronic activity
near a pair of charm mesons with small relative momentum.
They found that in events that
include a charm-meson pair with $k<300$~MeV, there are typically
two or three additional hadrons whose relative momentum with respect to 
one of the charm mesons satisfies $k_h < 100$~MeV.
Less than 10\% of the events have no such additional hadrons.
Since the estimate in Eq.~(\ref{sigma-est}) can accommodate the 
experimental lower bound on the prompt cross section for $X(3872)$
only if $k_\textrm{max} > 300$~MeV,
this level of hadronic activity is significant.

However hadronic activity near the molecule does not necessarily 
invalidate the use of the Migdal-Watson theorem.
The interaction between a generic low-momentum hadron and a 
constituent of a loosely-bound S-wave molecule is much weaker than 
the interaction between the two constituents.
For relative momentum $k$ in the range $\gamma_X < k < m_\pi$,
the interaction between the constituents is so strong 
that it saturates the unitarity bound.
For generic hadrons, the interaction strength may be close to 
the unitarity bound for $k \sim m_\pi$, but it does not increase at lower $k$.
An exception is a pion and $D$ meson with relative momentum 
of about 40~MeV, which have a P-wave resonance through the $D^*$.
With this exception, it is plausible that the effects of 
low-momentum hadrons can be treated as perturbations to the 
interactions between the constituents of the molecule.

\section{Discussion}
\label{sec:conc}

The deuteron and the $X(3872)$ 
(provided its quantum numbers are $J^{PC}=1^{++}$)
are manifestations of loosely-bound S-wave hadronic molecules.
As such, they have universal properties that are completely 
determined by their binding energies.
There have been several attempts in the literature to calculate 
their production rates based on the predictions of hadronic event 
generators for the production rates of their constituents. 
In the coalescence model, the production rate of the molecule 
is the production rate of a pair of 
its constituents integrated over the relative momentum 
up to $p_0$, as in Eq.~(\ref{sigma-cm}).
In the event-generator model defined in Section~\ref{sec:EvGen}, 
the production rate for the pair of 
constituents is also multiplied by a normalizing factor $K$,
as in Eq.~(\ref{sigma-egm}).  
It should be emphasized that these are purely phenomenological models.
The closest thing to a rigorous justification is the relation 
between the cross section for a loosely-bound S-wave molecule and the 
integrated cross section for its constituents in Eq.~(\ref{sigma-rig}).

In the spirit of hadronic event generators, 
the coalescence parameter $p_0$ and the normalizing factor $K$ 
should be treated as phenomenological parameters to be determined 
by experiment.
Theoretical estimates of these parameters, such as Eq.~(\ref{sigma-est})
with $k_\textrm{max} = 1/r_t$, can only provide order-of-magnitude
estimates.  Quantitative predictions using the event-generator model
require the determination of $p_0$ and $K$ from data.
The normalizing factor $K$ is necessary, because data on 
pairs of constituents are generally not among the inputs used to 
tune the event generator.  Since the naive cross section is 
proportional to $p_0^3$, the inclusive cross section for production 
of a molecule is sensitive only to $K p_0^3$.  One way to determine 
$K$ is from separate measurements of the production rate of a pair of
constituents.  For example, in 
Refs.~\cite{Bignamini:2009sk,Artoisenet:2009wk}, the cross section 
for $D^0 D^{*-}$ was used to determine the normalizing factor 
$K_{D^* \bar D}$ for the $X(3872)$.
Another way to determine $K$ is from separate measurements 
of both the molecule and the molecule plus its antiparticle.  
For example, CLEO data on inclusive $\bar d$ and
inclusive $\bar d + d$ was used in Section~\ref{sec:CLEO}
to estimate the normalizing factor $K_{np}$ for the deuteron.

We applied these considerations to the production of the anti-deuteron 
in $\Upsilon$ decays, confronting the predictions of the event-generator 
model with measurements by the CLEO Collaboration.
The measurement of the direct branching fraction for inclusive $\bar d$
in Eq.~(\ref{B:CLEO}) determines the combination $K_{np} p_0^3$.
The estimate of the branching ratio for inclusive $\bar d + d$
and inclusive $\bar d$ in Eq.~(\ref{BB:CLEO}) can then be used to
obtain separate estimates for $K_{np}$ and $p_0^3$.
These estimates suggest that the inclusive production rate for 
$\bar n \bar p$ 
is overestimated by Pythia, perhaps by an order of magnitude, 
and that the standard coalescence parameter $p_0$ underpredicts the
probability for formation of a $\bar d$ from $\bar n$ and $\bar p$ 
with small relative momentum.

We discussed the proposed upper bound on the production rate 
of a loosely-bound S-wave molecule that was derived
in Ref.~\cite{Bignamini:2009sk}.  If applied to the prompt 
cross section for the $X(3872)$ at the Tevatron,
the upper bound is more than two orders of magnitude smaller 
than the observed cross section,
leading the authors of Ref.~\cite{Bignamini:2009sk} to conclude 
that the $X(3872)$ can not be a loosely-bound molecule.
The analogous upper bound for the deuteron is given by 
Eq.~(\ref{sigma-ub}) with $k_\textrm{max} = k_0 + \Delta k$, 
where $k_0 = \gamma_d$ and $\Delta k = \gamma_d/2$
are estimates of the typical momentum and the momentum spread 
in the bound state. 
We demonstrated that the upper bound with this prescription for 
$k_\textrm{max}$ is invalid both on phenomenological and theoretical grounds.
The phenomenological grounds are that the prescription 
$k_\textrm{max}= 1.5~\gamma_d \approx 70$~MeV is smaller than
the standard coalescence parameter $p_0 = 80$~MeV.
This indicates that the anti-deuteron cross sections used to 
determine $p_0$ must exceed the proposed upper bound.
The prescription for $k_\textrm{max}$ in Ref.~\cite{Bignamini:2009sk}
is not only arbitrary, but it is based on the inappropriate use of
the minimal uncertainty principle for a Gaussian 
wavefunction to estimate $\Delta k$.  
A weakly-bound S-wave molecule maximizes the uncertainty, 
because $\langle k^2 \rangle = \infty$.
Thus the prescription in Ref.~\cite{Bignamini:2009sk}
underestimates the value of $k_\textrm{max}$
required for Eq.~(\ref{sigma-ub}) to be an upper bound on the cross section.
Since the naive $np$ cross section scales like $k_\textrm{max}^3$,
the upper bound on $\sigma[d]$ is underestimated by a much larger factor.
Similarly, the upper bound on the prompt cross section 
for the $X(3872)$ at the Tevatron is underestimated by 
more than an order of magnitude.
We conclude that there is no clear conflict between the observed 
cross section for $X(3872)$ and the interpretation 
of the $X(3872)$ as a loosely-bound charm-meson molecule.

The event-generator model should give the most accurate 
predictions for experiments that are the most similar to the ones used 
to determine the parameters $K$ and $p_0$.
Predictions for the LHC using parameters determined at the Tevatron
should be particularly accurate.
For the antideuteron, the inclusive differential cross section 
$d\sigma/dy$ at central rapidity $y=0$ has been measured by the 
E735 collaboration \cite{Alexopoulos:2000jk}.
This measurement can be used to determine $K_{np} p_0^3$,
which can then be used to predict cross sections for anti-deuteron
production at the LHC.
For the $X(3872)$, the prompt cross section in Eq.~(\ref{sigma-CDF}),
which was obtained from CDF measurements at the Tevatron,
can be used to determine 
$K_{D^* \bar D}~p_0^3~\textrm{Br}[X \to J/\psi\, \pi^+ \pi^-]$.
This combination can then be used to predict the 
prompt production rate of $X \to J/\psi\, \pi^+ \pi^-$ at the LHC.

One of the drawbacks of the event-generator model is the enormous 
number of events that must be generated to get reasonable 
statistics on the production rate of a pair of constituents 
with small relative momenta.  In the case of the $X(3872)$, 
there is a more efficient way to calculate the production rate.
The production of a charm meson pair with small relative momentum 
requires the creation of a $c \bar c$ pair with small relative momentum.
In the NRQCD factorization formalism, the production of the 
charm meson pair can be expressed as the sum of products of 
parton cross sections for the creation of the $c \bar c$ pair
and NRQCD matrix elements for the formation of the 
charm mesons~\cite{Braaten:2004jg}.
At leading order in $\alpha_s$, three of the four S-wave color/spin
$c \bar c$ channels have cross sections that are suppressed 
at large transverse momentum $p_T$ by at least a factor of 
$m_c^2/p_T^2$.  The $c \bar c$ channel that is not suppressed 
at leading order in $\alpha_s$ is color-octet $^3S_1$.  
Thus the simplest NRQCD factorization formula that can approximate
the predictions of the event-generator model is to keep only 
the color-octet $^3S_1$ term in the differential cross section:
%-----------------
\begin{equation}
d\sigma[X(3872)] = 
d\hat \sigma[c \bar c_8(^3S_1)]~\langle {\cal O}_8^X(^3S_1) \rangle. 
\label{sigma-NRQCD}
\end{equation}
%-----------------
The multiplicative constant $\langle {\cal O}_8^X(^3S_1) \rangle$
plays the same role as $K_{D^* \bar D} p_0^3$ in the 
event-generator model. In Ref.~\cite{Artoisenet:2009wk},
the combination 
$\langle {\cal O}_8^X(^3S_1) \rangle
	~\textrm{Br}[X \to J/\psi\, \pi^+ \pi^-]$
was determined from the prompt cross section at the Tevatron
in Eq.~(\ref{sigma-CDF}) and then used to predict the 
differential cross section of $X \to J/\psi\, \pi^+ \pi^-$ 
in various experiments at the LHC.
Similar results could presumably be obtained using the 
event-generator model, but the enormous number of Monte Carlo events
that would have to be generated makes it impractical.
The NRQCD factorization approach is much more efficient,
because the parton differential cross section 
$d\hat \sigma[c \bar c_8(^3S_1)]$
at leading order in $\alpha_s$ is known analytically.

\begin{acknowledgments}
This research was supported in part by the Department of Energy
under grant DE-FG02-91-ER40690. 

\end{acknowledgments}

%\newpage


\end{document}